# Cherenkov and scintillation light separation in organic liquid scintillators

J. Caravaca[1,2], F. B. Descamps[1,2], B. J. Land[1,2], M. Yeh[3], G. D. Orebi Gann[1,2,a]

[1] University of California, Berkeley, CA 94720-7300, USA
[2] Lawrence Berkeley National Laboratory, Berkeley, CA 94720-8153, USA
[3] Brookhaven National Laboratory, Upton, NY 11973-500, USA

**Abstract** The CHErenkov/Scintillation Separation experiment (CHESS) has been used to demonstrate the separation of Cherenkov and scintillation light in both linear alkylbenzene (LAB) and LAB with 2 g/L of PPO as a fluor (LAB/PPO). This is the first successful demonstration of Cherenkov light detection from the more challenging LAB/PPO cocktail and improves on previous results for LAB. A time resolution of $338 \pm 12$ ps FWHM results in an efficiency for identifying Cherenkov photons in LAB/PPO of $70 \pm 3\%$ and $63 \pm 8\%$ for time- and charge-based separation, respectively, with scintillation contamination of $36 \pm 5\%$ and $38 \pm 4\%$. LAB/PPO data is consistent with a rise time of $\tau_r = 0.72 \pm 0.33$ ns.

## 1 Introduction

The ability to separate Cherenkov and scintillation light in liquid scintillator (LS) detectors would enable a new generation of optical detectors that could achieve directional reconstruction at low energy thresholds, with powerful particle identification and event discrimination capabilities [1]. Such a detector would have applications across particle, nuclear, and astrophysics including a next-generation search for neutrinoless double beta decay, unprecedented sensitivity to solar neutrinos, proton decay, and sensitivity to the neutrino mass hierarchy and CP violation if deployed in a neutrino beam [1,2].

Cherenkov light is emitted at the picosecond scale after ionization, while typical time constants of organic liquid scintillators are at the nanosecond level; signal separation via photon arrival time is thus a theoretical possibility. Separation by photon density is feasible if an excess of Cherenkov photons can be observed above the isotropic scintillation background. Recent work [3] has demonstrated time separation in pure linear alkylbenzene (LAB). For experimental use LAB is most often deployed in combination with a fluor such as PPO to improve light yield by an order of magnitude. This substantially increases the physics reach of the experiment, but makes separation more challenging due to both significantly enhanced scintillation yield, which can swamp the Cherenkov component, and faster timing. Sub-ns time resolution becomes critical to achieve good separation. This can be achieved using fast PMTs, or new micro-channel based technology [4].

This article reports the first results from the CHErenkov Scintillation Separation experiment (CHESS) on Cherenkov / scintillation signal separation in pure LAB and LAB with 2 g/L of PPO (LAB/PPO). Further details of CHESS are described in [5], including the detector setup, calibrations, and observation of Cherenkov rings in water data, as well as Monte Carlo-based predictions for LS targets. This article contains the first results from LS data. Section 2 describes the experimental apparatus, Sect. 3 introduces the data taking procedure and analysis. The results for both target materials are presented in Sect. 4 and conclusions can be found in Sect. 5.

## 2 The CHESS experiment

Details of the CHESS experiment are described in [5]. This section provides an overview of the apparatus, along with calibrations and simulations relevant to LS data.

### 2.1 The apparatus

CHESS is designed to achieve sub-ns time resolution in photon detection using ultra-fast Hamamatsu H11934 photomultiplier tubes (PMTs) [6] with a transit time spread (TTS) of

[a] e-mail: gorebigann@lbl.gov

Springer

300 ps FWHM, read out with a 5 GHz CAEN V1742 digitizer [7]. 12 PMTs are positioned in a cross shape below a cylindrical UV-transmitting acrylic (UVT) target vessel (Fig. 1). A UVT block optically coupled between target and PMTs acts as an optical propagation medium. The primary source is through-going cosmic muons. Two muon tags placed above and below the setup provide a coincidence signal used to select muons within 6° from vertical, in order to produce a Cherenkov ring in a known location and scintillation light across the PMT array. A vertical hollow beneath the target prevents cosmic muons producing Cherenkov light in the propagation medium. The setup is surrounded by four Eljen [8] scintillator panels with $4\pi$ coverage of cosmic muons traversing the target. Three are used to veto shower events, which do not produce a clear Cherenkov ring. The fourth is beneath the target and, in combination with the muon tags, provides a triple coincidence for single-muon events. The majority of muon events are high energy and produce Cherenkov light at the maximum Cherenkov angle (48° in LAB; 41° in water). The target height and the spacing between target and PMTs are optimized such that the Cherenkov ring falls on a particular radial set of PMTs. The observed Cherenkov ring has a width defined by the target height, and falls at a position defined by the target-PMT separation. For example, in LAB the 48° and 6.5-cm target-PMT spacing results in an inner radius for the Cherenkov ring of approximately 7.2 cm (there is some shadowing caused by the vertical hollow, which is modeled in simulations), while the additional 3-cm target height puts the outer edge at 10.6 cm. This covers the outermost ring of PMTs. The target height is designed such that the width of the Cherenkov ring is approximately equal to the size of the PMT front face, to maximize photon detection efficiency. Isotropic scintillation light is detected across the entire PMT array. As a result, for a LS target the inner PMTs are hit only by scintillation photons and the outer PMTs by Cherenkov and scintillation photons (Fig. 1).

### 2.2 Monte Carlo simulation

In order to optimize the detector configuration and interpret results, a complete Monte Carlo (MC) simulation has been built using the RAT-PAC toolkit [9]. Geometry, material properties, PMT response, and DAQ effects are taken into account as described in [5]. A custom cosmic muon generator has been written following the Gaisser modified zenith and energy distribution [10]. Particle propagation and medium ionization is modeled by Geant4 [11], with the exception of optical photons, which are handled by the RAT-PAC custom absorption and reemission model. The time profile of scintillation light is modeled according to the triple exponential decay described in [12], with the addition of a single exponential rise time. LS properties are taken from [3,13–15].

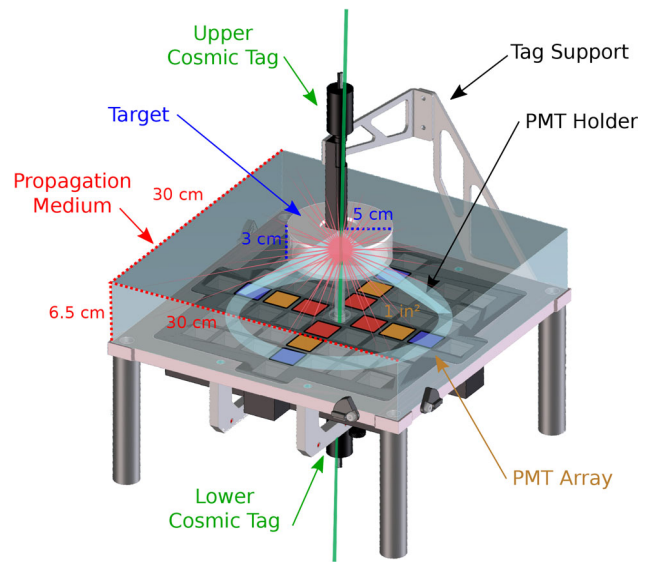

**Fig. 1** CHESS schematic. The PMT array can hold up to 53 PMTs; the 12 slots occupied for this study are color coded by radius: red and orange for those hit primarily by scintillation photons, and blue for those that detect the Cherenkov ring

## 3 Data taking and analysis

Cosmic muon events are collected with water, LAB, and LAB/PPO targets by triggering acquisition when the lower muon tag crosses a pre-defined threshold, set well above the single photoelectron (SPE) level to avoid triggering on PMT dark noise. Coincidence with the upper tag is required offline. A coincidence rate of $\sim 4\ \mu$/day is expected. Four weeks of data is taken with each target to ensure at least 100 events.

### 3.1 Calibration

The PMT response is characterized using calibration data prior to deployment of a LS target [5]. The PMT gain is measured using Cherenkov light from a $^{90}$Sr beta source with a water target, in order to provide a population of SPE hits. The mean number of PE per event is 2.6, equivalent to 0.22 PE per PMT per event on average, confirming that this source provides light at the SPE level. SPE charge distributions are collected for each PMT and fit with a Gaussian as an approximate model. The impact of the choice of model on the analysis is negligible. The mean of the Gaussian is taken as a measure of PMT gain, which is used to estimate the number of PE based on the charge of a hit PMT. The extracted gains were monitored periodically and determined to be very stable. The uncertainty on the gain was propagated as a systematic in the analysis, although the effect on the final result was determined to be negligible.

In order to accurately model hit times in the MC, the shape of individual PMT pulses must be understood. PMT pulse

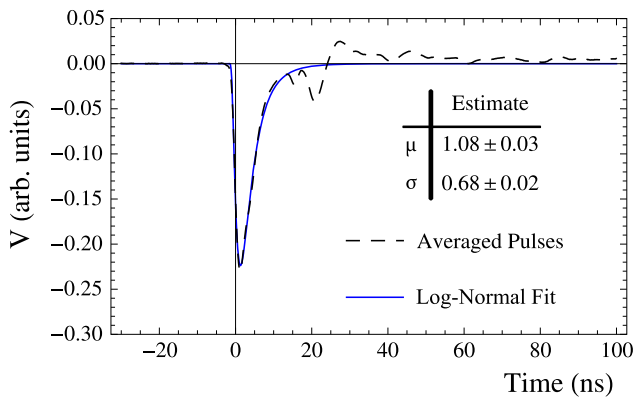
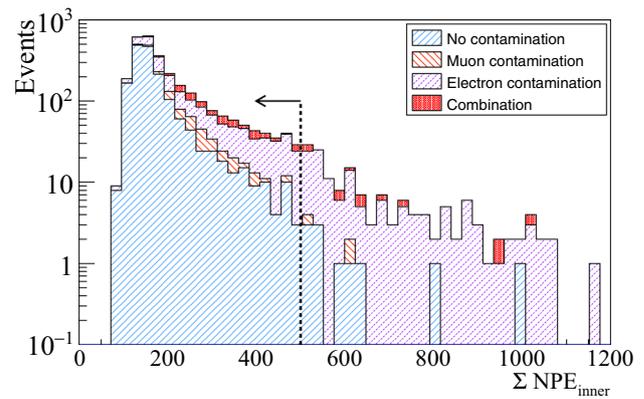

**Fig. 2** PMT pulse shape characterization. Log-normal fit to the average pulse shape across all PMTs. The pulse shape is normalized to unit area, hence voltage is reported in arbitrary units. The fit parameters reported are the mean and sigma of the log normal form. Taken from [5]

**Fig. 3** Monte Carlo simulation of the summed PE distribution on the inner PMT group due to cosmic muon events with perfect rings (turquoise, right diagonal hatching), muon contamination (orange, left diagonal hatching), electron contamination (purple, right diagonal dashed hatching), and the total (red, solid) for an LAB/PPO target. The vertical black dashed line represents the chosen cut value, with the arrow to illustrate the acceptance region. Taken from [5]

waveforms were collected for events close to the mean of the SPE distribution and an average was found across all PMTs. This was fit with a log normal distribution (Fig. 2), which is used to model pulse shapes in the MC. Further detail on this calibration can be found in [5].

Per-channel time delays due to electronic delays, cable delays, and the PMT transit time are measured using an LED, and cross-checked using Cherenkov light produced by cosmic muons passing through the propagation medium. The difference between the two sets of values is taken as a systematic uncertainty in the final analysis, as described in [5]. Values are on the order of hundreds of ps.

### 3.2 Event selection

Event selection criteria have been defined in order to obtain a clean sample of single cosmic muon events passing vertically through the target. Optimization of the selection is performed using a water-filled target and Monte Carlo simulation. A loose charge cut above the noise peak for the upper and lower muon tags and the panel immediately beneath the setup ensures coincidence events. To remove cosmic shower events, a detected charge consistent with electronic noise is required in the other 3 panels.

Large pulses from the PMT array induce crosstalk in the muon tag that occasionally trigger acquisition. This instrumental background can be identified by the characteristic oscillatory behavior of the resulting signals in the muon tags. The number of low frequency harmonic modes is calculated for the muon tag waveforms using a Fast Fourier Transform algorithm provided by ROOT [16]. Actual muon signals have a sizable low frequency component so events with a small low frequency component are removed.

Another undesirable background is caused by muons clipping the propagation medium, or producing secondary particles that can themselves generate photons in that medium.

Both create some Cherenkov light contamination across the PMT array, primarily on the innermost PMTs. These events have been modeled in simulation [5], and the expected number of PE from these events is seen to be appreciably greater even than the expected scintillation yield, due to the proximity of production in the propagation medium. A cut was designed to reject these events based on event topology: since the Cherenkov ring geometry is not expected to produce hits on the inner PMTs for either a water or LS target, the total number of estimated PEs on the inner PMTs was used to identify clean rings. Between 30 and 400 PE of Cherenkov photon contamination is expected on the inner PMTs for these events, whereas the expected scintillation yield on these PMTs is $\sim$ 30 PEs for LAB and $\sim$ 300 PEs for LAB/PPO. The summed charge on these PMTs can therefore be used to identify and remove these events. A threshold is applied to the total number of PEs across the 4 inner PMTs, $NPE_{inner}$, rejecting events that have $NPE_{inner} > 40$ for LAB and $NPE_{inner} > 500$ for LAB/PPO. The cut value was chosen based on simulations (Fig. 3) to reject events with high contamination due to secondary particles with a minimal impact on the efficiency.

### 3.3 Charge and time measurements

The signal from each PMT is digitized at 5 GHz and the resulting waveforms are analyzed to extract the hit time and charge. The charge is defined as the pedestal-corrected integral of the pulse in a 135 ns window, with the pedestal calculated on a per-waveform basis. Traces with a baseline fluctuation larger than 5 mV peak-to-valley in the pedestal region are not analyzed. An empty channel on the digitizer was used to monitor and correct for any intra-event baseline trends. The

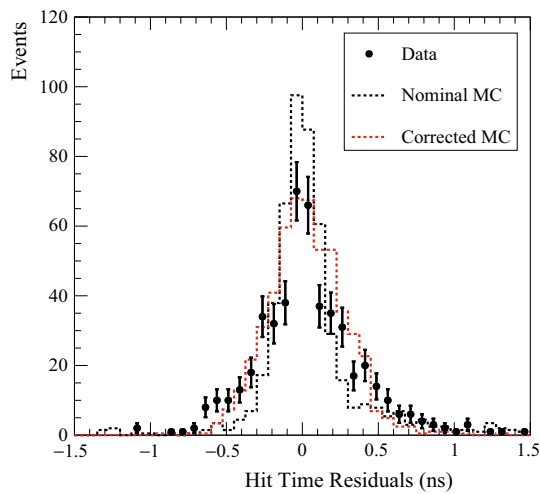

**Fig. 4** Distribution of hit-time residuals with respect t reconstructed event time for ring candidate events in water for middle PMTs, where the Cherenkov ring is expected. Data points are shown with statistical errors, with the Monte Carlo prediction overlaid (dashed lines). Taken from [5]

uncertainty on this correction was propagated through the extraction of PMT gains (Sect. 3.1).

The hit time is calculated using a fixed threshold approach, with a channel-dependent threshold set at 1/4 of the SPE peak. A simple algorithm reconstructs the event time as the median of the four most prompt hits. Averaging over four hits provides some stability against potential early noise hits and PMT TTS jitter while maintaining good time resolution since the Cherenkov yield is such that these prompt hits are due to Cherenkov light, whose time profile is very sharp. Hit-time residuals are calculated as the difference between the PMT hit time and the event time, corrected for photon time-of-flight and the measured per-channel time delays.

The time resolution of CHESS is determined from the hit-time residual distribution of Cherenkov events produced by cosmic muons with a water target [5]. The resolution is found to be $338 \pm 12$ ps FWHM.

A comparison of data to MC prediction for water show that the simulation slightly under-predicts the detector resolution, likely due to PMT-to-PMT variations in the PMT TTS and pulse shapes, or multi-photon effects (Fig. 4). This is corrected for in simulations for all targets by including an additional smearing on the hit-time residual distributions of $\sigma = 214$ ps [5].

## 4 Results

### 4.1 Separation in LAB

After applying event selection criteria, 117 ring candidates were identified in the LAB data set. An example event is shown in Fig. 5, while the average across the data set is shown in Fig. 6 for both time and charge. A clear ring structure can be observed in both cases.

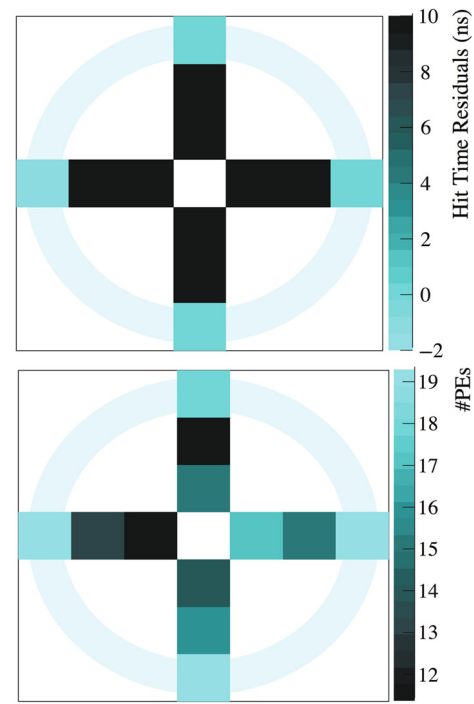

**Fig. 5** A single ring candidate event in LAB. (Top) Hit-time residual and (Bottom) number of PE versus PMT position. No hit selection is applied here, i.e. this is prior to application of the $t_c$ timing cut

The hit-time residual distributions for LAB are shown in Fig. 7 individually for the three radial PMT groupings, for data and MC. The outer PMTs, where the Cherenkov ring is expected, register the earliest hits, while the distributions for the middle and inner groups are broader and peak later, consistent with scintillation light. Early features in the inner and middle PMT groups are primarily due to Cherenkov light contamination from secondary electrons. The MC reproduces data well, supporting the conclusion that true Cherenkov/scintillation separation has been observed. The first photon hits on the outer PMTs are due to fast Cherenkov photons, whereas middle and inner PMT hits are due to scintillation light. The ratio of hit counts between the two populations can thus be used to quantify the Cherenkov/scintillation separation. A time cut ($t < t_c$) is optimized in order to maximize separation. The efficiency of identifying Cherenkov hits is defined as the fraction of outer PMT hits that occur before $t_c$. The scintillation contamination is given by the fraction of hits occurring before $t_c$ that are due to scintillation i.e. hits with $t < t_c$ on the inner and middle PMTs. The analysis is performed using both LED- and muon-extracted time delays and the difference taken as a systematic uncertainty. Uncertainties in the time-of-flight calculation are also included in the systematic evaluation, although this is a subdominant effect. The optimal timing cut for LAB is $t_c = 0.4$ ns, yielding $83 \pm 2(\text{stat}) \pm 2(\text{syst})\%$ detection efficiency for Cherenkov hits, with contamination of $11 \pm 1(\text{stat}) \pm 0(\text{syst})\%$. Better separation might be achieved by eliminating the Cherenkov contamination in the inner and middle PMTs via improvements to the CHESS apparatus.

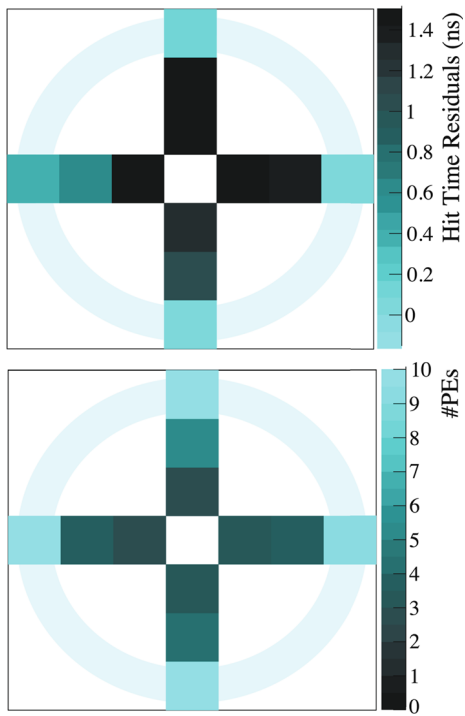

**Fig. 6** (Top) Average hit-time residuals and (Bottom) average number of PE versus PMT position for ring candidate events in LAB. No hit selection is applied here, i.e. this is prior to application of the $t_c$ timing cut

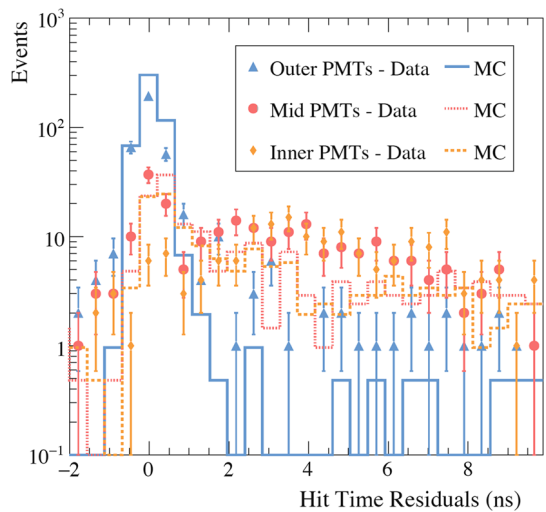

**Fig. 7** Hit-time residual distributions in LAB for data and MC

Charge separation can be achieved by taking the ratio of charge in a prompt window to that in the full event window for each hit PMT and comparing this for the different populations of hits. The result for LAB is shown in Fig. 8. The separation is defined in an analogous manner to that for time: a threshold is optimized to maximize the Cherenkov hit detection efficiency and minimize scintillation contamination. A threshold of $Q_{\text{ratio}} = 0.09$ gives a Cherenkov detection efficiency of $96 \pm 2(\text{stat}) \pm 0(\text{syst})\%$ with contamination of $6 \pm 3(\text{stat}) \pm 0(\text{syst})\%$.

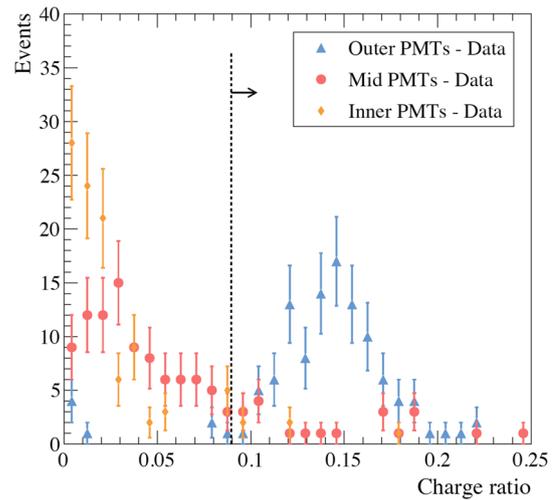

**Fig. 8** Ratio of charge in a prompt, 5 ns window to the total charge for each hit PMT for pure LAB

### 4.2 Separation in LAB/PPO

After applying event selection criteria, 103 ring candidates were identified in the LAB/PPO data set. The topology of a typical event, and the average across the data set, are shown in Fig. 9. Again a clear ring structure can be observed. Hit-time residual distributions are shown in Fig. 10, where the early Cherenkov hits on outer PMTs are clear. The separation is less distinct than for pure LAB, as expected. The MC shows good agreement with data. A cut at $t_c = 0.4$ ns yields Cherenkov detection efficiency of $70 \pm 3(\text{stat}) \pm 0(\text{syst})\%$ with contamination of $36 \pm 3(\text{stat}) \pm 4(\text{syst})\%$.

A scan of the LAB/PPO rise time, $\tau_r$, showed that the data is inconsistent with zero rise time. A scan of the rise time, propagating the systematic uncertainties as described in Sects. 3.1 and 3.3 and in more detail in [5], results in a value of $\tau_r = 0.72 \pm 0.33(\text{stat}) \pm 0.01(\text{syst})$ ns.

The charge ratio for LAB/PPO is shown in Fig. 11. The separation is more difficult in LAB/PPO due to higher scintillation yield. A threshold of $Q_{\text{ratio}} = 0.038$ gives a Cherenkov detection efficiency of $63 \pm 4(\text{stat}) \pm 7(\text{syst})\%$ with contamination of $38 \pm 3(\text{stat}) \pm 2(\text{syst})\%$.

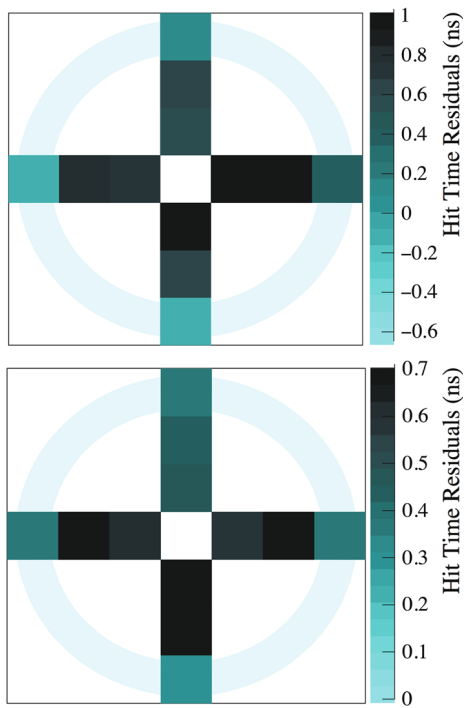

**Fig. 9** Hit-time residuals versus PMT position for (Top) a single ring candidate event and (Bottom) the average across all events in LAB/PPO. No hit selection is applied here, i.e. this is prior to application of the $t_c$ timing cut

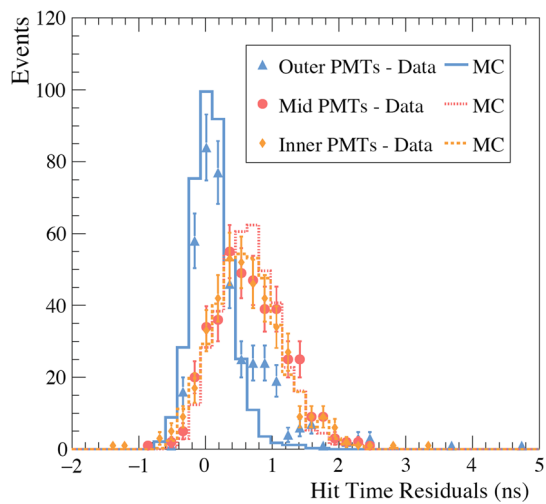

**Fig. 10** Hit-time residual distributions in LAB/PPO for data and MC

## 5 Conclusions

Successful time- and charge-based separation of Cherenkov and scintillation light has been achieved in both LAB and LAB/PPO with CHESS. A time resolution of $338 \pm 12$ ps FWHM yields a Cherenkov detection efficiency in LAB/PPO of $70 \pm 3\%$ and $63 \pm 8\%$ for time- and charge-based separation, respectively, with scintillation contamination of $36 \pm 5\%$ and $38 \pm 4\%$. In pure LAB the time- and charge-based Cherenkov efficiencies are $83 \pm 3\%$ and $96 \pm 2\%$, with contaminations of $11 \pm 1\%$ and $6 \pm 3\%$. The LAB/PPO data is inconsistent with zero rise time, and consistent with a value of $\tau_r = 0.72 \pm 0.33(\text{stat}) \pm 0.01(\text{syst})$ ns.

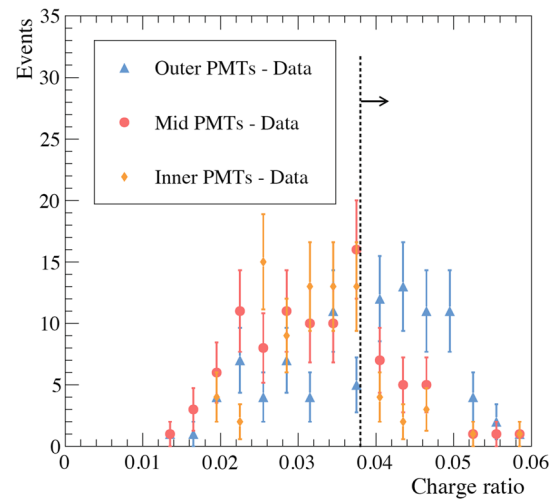

**Fig. 11** Ratio of charge in a prompt, 5 ns window to the total charge for each hit PMT for LAB/PPO

Demonstration of Cherenkov/scintillation separation at the PMT hit level opens new possibilities for optical detectors. Cherenkov rings have been successfully identified in both LAB and LAB/PPO. This capability in a larger detector would allow reconstruction of incident particle direction. Even without explicit Cherenkov ring topology, the quantitative charge- and time-based separation presented here would allow identification of Cherenkov PMT hits within an event, which could then be used to reconstruct particle direction. Thus these results demonstrate the required capabilities for directional reconstruction in a large-scale detector. Realizing this result in a large-scale scintillator detector such as such as SNO+ [17] or a future large-scale LS detector [18] would significantly improve sensitivity to neutrinoless double beta decay by allowing rejection of the dominant $^8$B solar neutrino background, as well as improving sensitivity to a range of physics topics including solar and supernova neutrinos. Studies presented in [1] discuss the potential impact of this separation on the sensitivity of a large detector to a broad range of physics goals. This study assumed a water-based LS target with a conservatively low light yield, however, the new results presented in this paper demonstrate the potential to identify the Cherenkov component even in a pure LS target, thus allowing directional sensitivity in a high light yield detector. Studies in [19,20] discuss how this separation could be used to extract particle direction in a pure LS detector.

**Acknowledgements** This work was supported by the Laboratory Directed Research and Development Program of Lawrence Berkeley

National Laboratory under U.S. Department of Energy Contract No. DEAC02- 05CH11231. The work conducted at Brookhaven National Laboratory was supported by the U.S. Department of Energy under contract DE-AC02-98CH10886. The authors would like to thank the SNO+ collaboration for providing data on the optical properties of LAB/PPO, including the light yield, absorption and reemission spectra, and refractive index.